\documentclass[pdflatex,sn-nature]{sn-jnl}% Style for submissions to Nature Portfolio journals
\usepackage{graphicx}%
\usepackage{multirow}%
\usepackage{amsmath,amssymb,amsfonts}%
\usepackage{amsthm}%
\usepackage{mathrsfs}%
\usepackage[title]{appendix}%
\usepackage{xcolor}%
\usepackage{textcomp}%
\usepackage{manyfoot}%
\usepackage{booktabs}%
\usepackage{algorithm}%
\usepackage{algorithmicx}%
\usepackage{algpseudocode}%
\usepackage{listings}%
\usepackage{subcaption} % Required for subfigures
\usepackage{graphicx}
\theoremstyle{thmstyleone}%
\theoremstyle{thmstyletwo}%

\theoremstyle{thmstylethree}%

\newcommand\blfootnote[1]{%
  \begingroup
  \renewcommand\thefootnote{}\footnote{#1}%
  \addtocounter{footnote}{-1}%
  \endgroup
}
\begin{document}

\title[Article Title]{From Model Uncertainty to Human Attention: Localization-Aware Visual Cues for Scalable Annotation Review}

% \author*[1,2]{\fnm{First} \sur{Author}}\email{iauthor@gmail.com}

\author[1]{\fnm{Moussa} \sur{Kassem Sbeyti$^\dagger$}}%\email{moussa.sbeyti@kit.edu}
% \equalcont{Equal contribution. Authors are permitted to list their name first.}

\author[2]{\fnm{Joshua} \sur{Holstein$^\dagger$}}%\email{joshua.holstein@kit.edu}
% \equalcont{Equal contribution. Authors are permitted to list their name first.}
\author[2]{\fnm{Philipp} \sur{Spitzer}}%\email{philipp.spitzer@kit.edu}
\author[1]{\fnm{Nadja} \sur{Klein}}%\email{nadja.klein@kit.edu}
\author[2]{\fnm{Gerhard} \sur{Satzger}}%\email{gerhard.satzger@kit.edu}

\affil[1]{\orgdiv{Scientific Computing Center}, \orgname{Karlsruhe Institute of Technology}}
\affil[2]{\orgdiv{Institute for Information Systems}, \orgname{Karlsruhe Institute of Technology}}


\abstract{
High-quality labeled data is essential for training robust machine learning models, yet obtaining annotations at scale remains expensive. AI-assisted annotation has therefore become standard in large-scale labeling workflows. However, in tasks where model predictions carry two independent components, a class label and spatial boundaries, a model may classify an object with high confidence while mislocalizing it. Existing AI-assisted workflows offer annotators no signal about where spatial errors are most likely. Without such guidance, humans may systematically underinspect subtly misplaced boxes. 
We address this by studying the effect of visualizing spatial uncertainty via a purpose-built interface. In a controlled study with 120 participants, those receiving uncertainty cues achieve higher label quality while being faster overall. A box-level analysis confirms that the cues redirect annotator effort toward high-uncertainty predictions and away from well-localized boxes. These findings establish localization uncertainty as a lever to improve human-in-the-loop annotation. Code is available at \url{https://mos-ks.github.io/MUHA/}.
\blfootnote{$^\dagger$Equal contribution. Authors are permitted to list their name first.}
}




\maketitle

\section*{Main}
\label{sec:main}
Machine learning models are increasingly deployed in human-in-the-loop systems, where a human reviewer must verify model outputs before they inform consequential decisions~\cite{senoner2024explainable}. The reliability of such systems depends on the reviewer's ability to detect model errors~\cite{senoner2024explainable}. Empirical work shows this assumption regularly fails as humans tend to overrely on model predictions and miss errors that are not explicitly flagged~\cite{vaccaro2024combinations}. To address this, uncertainty estimates alongside predictions have been shown to direct human attention toward instances where predictions are more likely to be wrong~\cite{Holstein2025}, thereby improving the overall performance~\cite{bhatt2021uncertainty,vaccaro2024combinations}. However, existing work has focused almost exclusively on classification settings, in which uncertainty can be communicated via a single value associated with a predicted label~\cite{guo2017calibration, lakshminarayanan2017simple}.

In spatial domains such as object detection, however, a prediction has two independent components: the object class with a corresponding score and the object's spatial boundaries. A detector can be highly confident in a prediction while mislocalizing it, meaning that a high-confidence prediction carries no information about whether its boundaries are correctly placed~\cite{choi2021active,kassem2024cost}. Deploying such miscalibrated object detectors in the real-world carries a significant risk. A detector is as good as its training data. Errors in ground truth labels enter the training loop and compound across retraining cycles, making label quality critical for model performance. Yet the long tail of rare but critical events in high-stakes domains can only be covered through volume, making manual labeling alone prohibitively expensive. To meet these demands, crowd-labeling platforms have emerged, coordinating many workers who collectively spend millions of hours labeling images~\cite{lin2014microsoft,russakovsky2015imagenet}. To reduce the manual labeling efforts and time, AI-assisted labeling has become the standard workflow, in which an AI model predicts initial labels for human annotators to review and correct. 

This labeling workflow directly inherits the asymmetry described above: while reviewers receive a model confidence score for each predicted class, no equivalent signal exists for the spatial precision of the predicted boundaries, leaving them without guidance on where localization errors are most likely to occur. Misclassifications, such as a cyclist labeled as a tram, also produce a categorical mismatch that is visually apparent and relatively easy to recognize. A mislocalized boundary lacks this signal, and reviewers need to default to heuristic attention allocation, thereby systematically underinspecting the spatial dimensions of predictions. Such errors are widespread even in established benchmarks~\cite{northcutt2021confident}, with well-documented consequences for model generalization~\cite{song2022learning}, and in safety-critical domains such as autonomous driving or healthcare, they translate directly into performance degradation~\cite{nguyen2023efficient,sbeyti2025building} and inference failures~\cite{saeeda2025data}.

Existing approaches to improving annotation quality operate mostly at the level of which images to review~\cite{zhan2022comparative}. Localization uncertainty has proven an effective proxy for this selection~\cite{kao2018localization,choi2021active,yang2024plug}. Of particular relevance is aleatoric uncertainty, which captures irreducible noise in the observed data (in contrast to epistemic uncertainty arising from model uncertainty)~\cite{kendall2017uncertainties, hullermeier2021aleatoric}. Aleatoric localization uncertainty is found to correlate with occlusion, box accuracy, and image quality~\cite{kassem2023overcoming}. Since it is irreducible by the model or by collecting more data, improving the quality of ground-truth labels is the only viable means of limiting its effect on model training. Yet existing annotation interfaces focus primarily on speed and cognitive load: drawing tools snap boxes to edges~\cite{williams2024snapper}, replace corner-dragging with extreme clicks~\cite{papadopoulos2017extreme}, or accept free-hand sketches via feedback from large language models~\cite{li2025s}, while model-assisted workflows pre-propose boxes for human correction~\cite{adhikari2021iterative} and interface design reduces burden by surfacing guidelines and constraining invalid selections~\cite{henley2024supporting}. None of these exploits model-estimated localization uncertainty as a source of attentional guidance, leaving a systematic source of label degradation intact even in otherwise well-designed pipelines.

\begin{figure}[!ht]
    \centering
    \includegraphics[width=\linewidth]{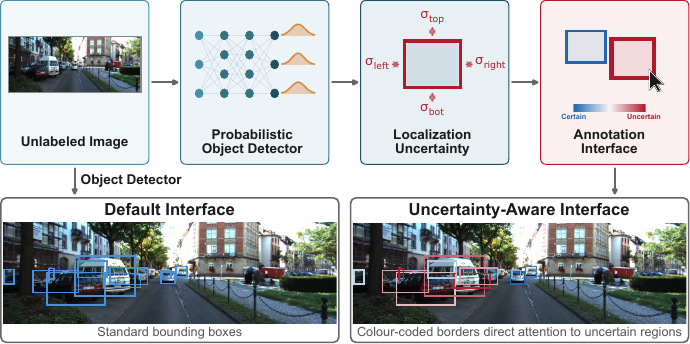}
    \caption{\textbf{Overview.} In the default interface, annotators review bounding boxes produced by a standard detector without guidance. In the uncertainty-aware interface, a probabilistic object detector produces bounding box predictions with per-coordinate aleatoric localization uncertainty. Bounding box borders are color-coded from blue (certain) to red (uncertain), directing attention toward spatially unreliable predictions.}
    \label{fig:framework}
\end{figure}

We close this gap by studying the effect of communicated model-based spatial uncertainty on annotators. A purpose-built interface visualizes aleatoric spatial uncertainty of labels as a color-coded visual cue. These cues direct annotator attention toward bounding boxes with unreliable boundaries (see Figure~\ref{fig:framework}). In a randomized controlled study, 120 participants reviewed model-generated labels in 1,800 annotation trials on autonomous driving images. We observe that participants receiving uncertainty cues are 7.2\% faster overall and achieve higher label quality by 0.70 mean Intersection over Union (mIoU) \%-points and that these benefits increase with higher image difficulty. An analysis of individual cognitive load differences suggests that the interface succeeds not by reducing overall cognitive demand, but by directing attention toward the boxes that most require human correction. These findings establish localization uncertainty as a practical lever for improving human-in-the-loop image annotation, specifically where model predictions carry a spatial component that class confidence scores alone do not capture. In high-stakes domains such as autonomous driving, where large volumes of labeled data must be continuously generated and updated for model redeployment, labeling errors directly propagate into safety-critical inference failures. By simultaneously improving label quality and reducing annotation time, our approach offers a scalable path toward higher-quality training data at lower cost, enabling more labels to be reviewed within the same budget and accelerating model deployment.

\section*{Results}
\label{sec:results}

\subsection*{Localization Uncertainty Improves Annotation Quality and Efficiency}
Our primary investigation assessed whether providing localization uncertainty improves the quality and efficiency of human-in-the-loop annotations. We conducted a randomized controlled experiment with 120 participants ($N = 1{,}800$ trials), comparing a standard interface (\textit{Baseline}) against our proposed uncertainty-aware interface (\textit{Uncertainty Visualization}). For a more granular analysis on how uncertainty communication supports human annotators, we define three image difficulty levels (Easy, Medium, Difficult) by binning the predicted detector uncertainty.
 
\begin{figure}[h!]
    \centering
    \includegraphics[width=\linewidth]{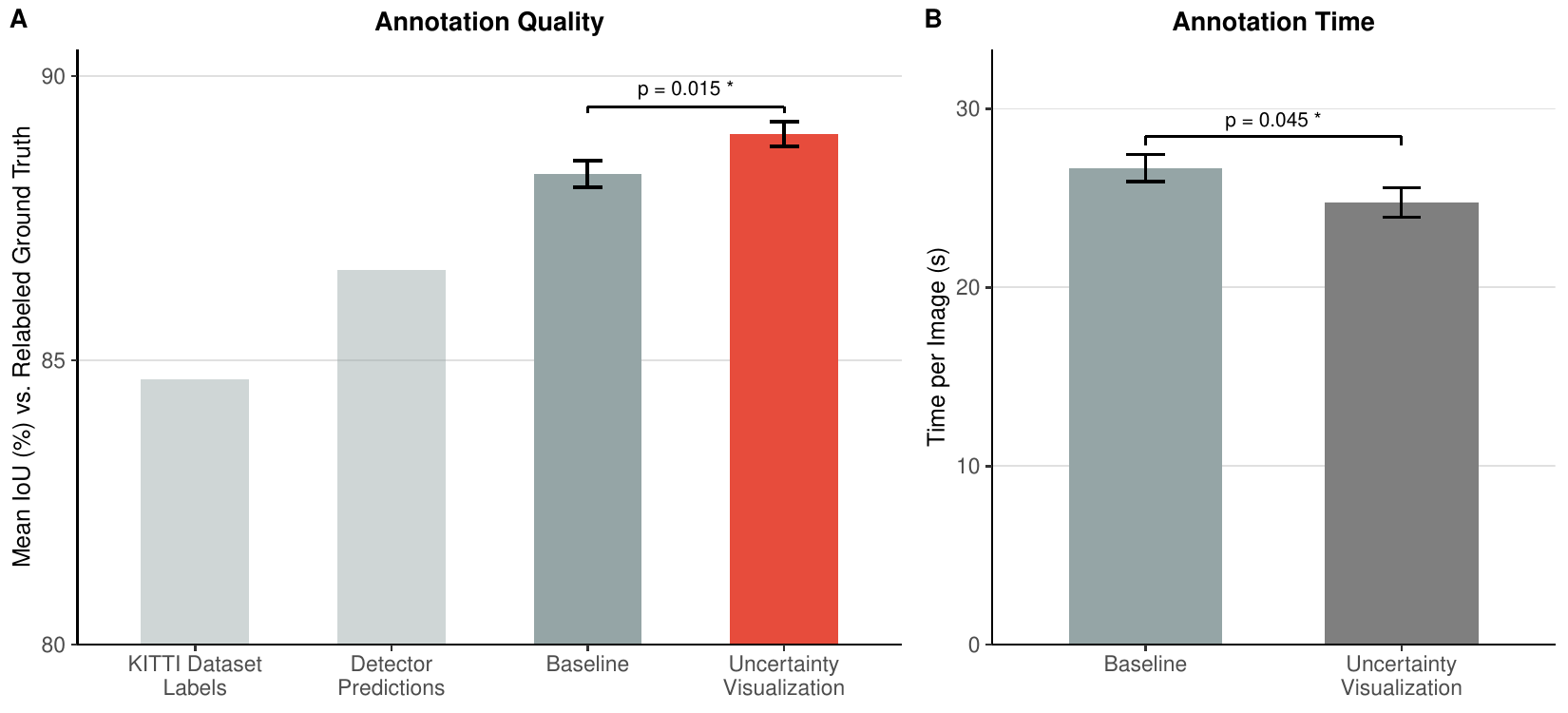}
    \caption{\textbf{Impact of Localization Uncertainty on Annotation Performance.} \textbf{(A)} Annotation Quality (mIoU) across the entire labeling pipeline. KITTI dataset labels and detector predictions serve as reference points, illustrating the quality baseline from which human annotators begin. Participants in the uncertainty visualization condition (red) achieve higher annotation quality than the baseline condition ($p=0.015$). \textbf{(B)} Annotation Efficiency: The treatment group is also significantly faster ($p=0.045$), demonstrating a dual benefit. Error bars represent the Standard Error of the Mean.}
    \label{fig:performance}
\end{figure}

\textbf{Annotation Quality.} As shown in Figure~\ref{fig:performance} (A), mIoU increases monotonically throughout the entire labeling pipeline, evaluated against the author-relabeled ground truth reported in \ref{sec:method}. The original KITTI labels score 84.7\%, reflecting the imprecision our relabeling effort corrected. Detector predictions then improve this to 86.9\%, baseline participants further improve to $88.28\% \pm 1.81$, and those receiving uncertainty cues reach $88.98\% \pm 1.69$, demonstrating that the human-AI team achieves complementary team performance, outperforming what either achieves in isolation~\cite{hemmer2025complementarity}. A one-tailed independent-samples $t$-test confirms the improvement from baseline to treatment as statistically significant ($t(117.4) = 2.19$, $p = 0.015$, $d = 0.4$)., indicating a medium-sized effect size, showing robust enhancement in annotation quality.

To verify the robustness of this result against individual annotator variability, we fit a Linear Mixed Effects (LME) model using Maximum Likelihood (ML) estimation, specifying treatment condition as a fixed effect and including random intercepts for participants to account for the repeated-measures structure of the data. This analysis confirms the significant positive main effect of the treatment ($\beta = 0.70$, $SE = 0.32$, $t = 2.20$, $p = 0.030$, two-tailed). The treatment effect remains consistent in sign and magnitude when winsorizing mIoU values and when adjusting for self-efficacy and task familiarity as covariates. When additionally controlling for initial model prediction quality, the effect attenuates to a non-significant trend ($\beta = 0.40$, $p = 0.136$), likely because this covariate absorbs variance on images with few objects where uncertainty cues offer little theoretical benefit. Yet, the treatment effect survives this check in the subset of uncertain images (medium$+$difficult) (Supplementary Table~S1; see also the Guidance Effect section below).

\textbf{Efficiency.} Quality improvement does not come at the cost of efficiency. The uncertainty-aware group is even more efficient (Figure~\ref{fig:performance}, B), reducing the average annotation time per image from $26.69$~s to $24.76$~s, a $7.2\%$ reduction. A one-tailed independent-samples $t$-test confirms this improvement as statistically significant ($t(117.4) = -1.714$, $p = 0.045$, $d = -0.313$).

To verify the robustness of this result, we fit a parallel LME model with treatment condition as a fixed effect and random intercepts for participants. The treatment effect is directionally consistent but does not reach significance at the two-tailed threshold ($\beta = -1.93$, $SE = 1.12$, $t = -1.73$, $p = 0.086$, two-tailed), indicating that the overall efficiency advantage is more modest than the quality gain. The efficiency benefit sharpens considerably on medium and difficulty images, where the LME confirms a significant effect ($\beta = -2.37$, $SE = 1.15$, $p = 0.042$, two-tailed).

\subsection*{The ``Attention Guidance Effect'': Benefits Scale with Image Difficulty}

We first examine where annotators direct their effort at the level of individual model predictions to understand the mechanism underlying the overall performance gains. For each original prediction box, we Hungarian-match it against the annotator's submitted boxes for that image and record whether the bounding box was changed at all (any IoU drop $>1\%$), relating it to the model's predicted localization uncertainty. We then examine whether the guidance effect on performance varies systematically across images of different difficulty levels.

\begin{figure}
    \centering
    \includegraphics[width=1\linewidth]{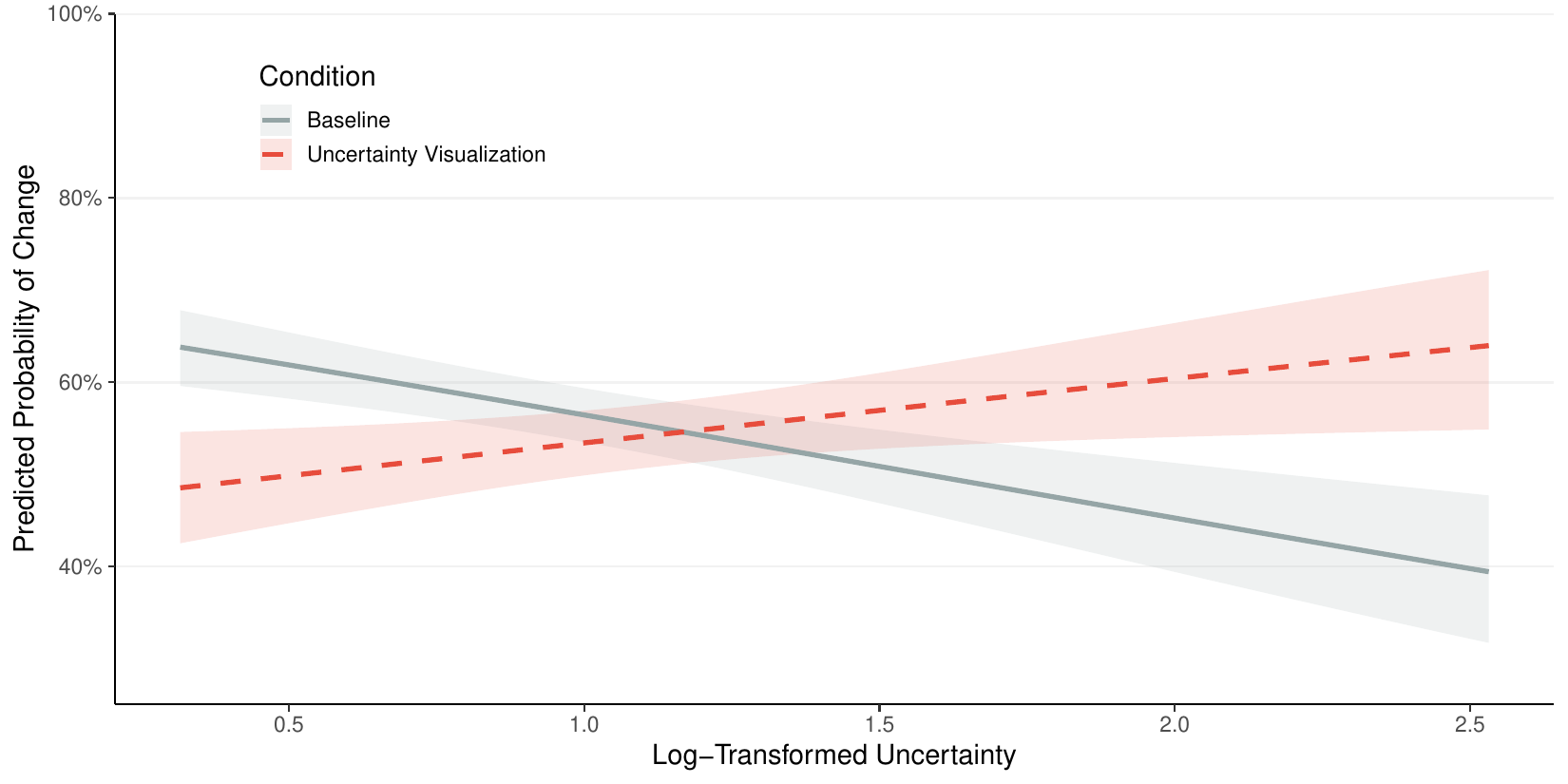}
    \caption{Guidance effect of uncertainty visualization on intervention behavior. Predicted probability of box modification as a function of log-transformed uncertainty. While the likelihood of manual intervention decreases at higher uncertainty levels in the Baseline condition, the Uncertainty Visualization condition significantly increases the probability of corrective action. Lines represent GEE model estimates; ribbons indicate 95\% Wald CIs.}
    \label{fig:pct-changed-albox}
\end{figure}

\textbf{Treatment successfully guides attention to uncertain boxes.} In the uncertainty condition, the share of edited boxes increases with the corresponding uncertainty, showing that annotators concentrate effort on precisely the predictions the model is least confident about (Figure~\ref{fig:pct-changed-albox}). In the baseline condition, this pattern reverses: without a guidance signal, annotators disproportionately adjust low-uncertainty predictions that already align well with the ground-truth, while leaving high-uncertainty boxes largely untouched, even though those are the ones that most require correction.

To formally test this attention-reallocation effect, we model the probability that a bounding box is edited ($N = 10{,}574$ observations, 120 annotators) using Generalized Estimating Equations (GEE) logistic regression with an exchangeable working correlation structure (assuming equal correlation between any two boxes from the same annotator) and sandwich-corrected standard errors, including experimental condition, log-transformed localization uncertainty, and their interaction as predictors. The results confirm the pattern in Figure~\ref{fig:pct-changed-albox}. The negative main effect of condition ($\beta = -0.858$, $p < .001$) indicates that treatment annotators are less likely to edit low-uncertainty boxes than those in the baseline. The negative main effect of uncertainty in the baseline ($\beta = -0.45$, $p < .001$) quantifies the declining edit rate without any guidance signal. The positive and significant interaction ($\beta = 0.736$, $p < .001$) captures the reversed slope in the uncertainty condition, where the net effect of uncertainty is positive ($0.736 - 0.45  = 0.286$), confirming that annotator effort shifts toward higher-uncertainty predictions. Together, these box-level results establish that the color-coded signal was behaviorally effective in redirecting attention where it is most needed.

\begin{figure}[t]
    \centering
    \includegraphics[width=\linewidth]{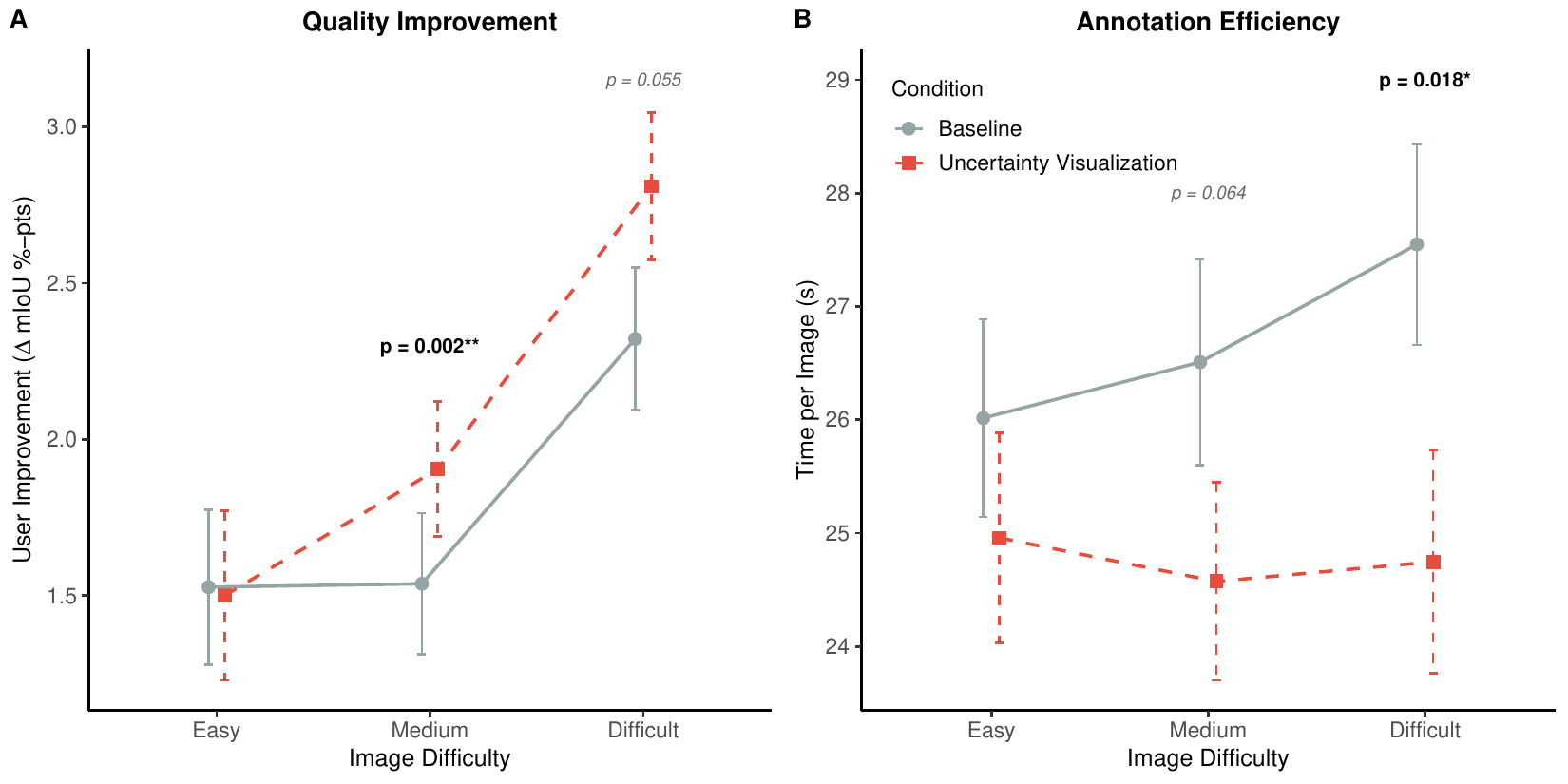}
    \caption{Efficacy by task difficulty. The benefits of the uncertainty visualization scale with image difficulty. \textbf{(A) Quality Improvement:} Measured as the change from the initial model prediction ($\Delta$ mIoU). \textbf{(B) Annotation Efficiency:} Measured in seconds per image. While performance and speed are similar on easy images, the treatment group shows significantly greater quality improvement on medium-difficulty images and significantly faster completion times on high-difficulty images, indicating effective attentional guidance.}
    \label{fig:difficulty}
\end{figure}

\textbf{The attentional guidance translates into performance gains that scale with image difficulty.} If uncertainty cues work by narrowing the effective visual search space, their benefit should be negligible where the model is already well-calibrated, and increase as images become more ambiguous and spatially demanding. Figure~\ref{fig:difficulty} confirms this hypothesis. On easy images, where model predictions are already well-localized and few boxes compete for attention, the treatment offers no significant advantage for either quality, i.e., effectiveness ($p = 0.227$, $d = 0.14$) or efficiency ($p = 0.20$, $d = -0.15$). As image difficulty increases, the benefits become more pronounced. For medium-difficulty images, the treatment group achieves significantly higher quality ($p = 0.002$, $d = 0.53$) and tends to be faster (${\sim}2.0$~s reduction, $p = 0.064$, $d = -0.28$). For the most challenging images, the treatment group shows a directional trend in quality ($p = 0.055$, $d = 0.29$) and is faster (${\sim}2.8$~s reduction, $p=0.018, d=-0.386$). This difficulty-scaling pattern is precisely what an attentional guidance account predicts. On images categorized as easy, annotators can identify well-localized boxes without help, so the cue is redundant; on harder images, where spatial errors are subtle and multiple boxes compete for inspection, the cue reduces the effective search space and directs effort toward the annotations most in need of correction, consistent with classic accounts of guided visual search~\cite{itti2001computational}.

To further test this interpretation, we conduct a secondary analysis restricting the sample to medium- and high-difficulty images, the conditions under which uncertainty cues are theoretically informative. In this subset, both effects sharpen. Annotation quality improves to $d = 0.49$ ($t(117.9) = 2.69$, $p = 0.004$), confirmed by LME ($\beta = 0.87$, $SE = 0.32$, $p = 0.008$), and the efficiency advantage likewise strengthens ($d = 0.37$, $t(117.7) = -2.04$, $p = 0.022$; LME: $\beta = -2.37$, $SE = 1.15$, $p = 0.042$). Importantly, when controlling for initial model prediction quality within this subset, the treatment effect maintains significance ($\beta = 0.55$, $SE = 0.27$, $p = 0.040$), ruling out the possibility that the benefit on harder images is merely a proxy for differences in starting annotation quality. Full robustness check results are reported in Supplementary Tables~S1--S2. Taken together, the box-level attention data and the difficulty-stratified image-level outcomes mutually reinforce a single account: uncertainty visualization works by redirecting annotator effort toward the spatial predictions that most require human correction, and this redirection produces meaningful gains precisely where the model's spatial outputs are least reliable.

% -----------------------------------------------------------------------------
\subsection*{Cognitive Load and Treatment Effects}
% -----------------------------------------------------------------------------
To examine whether individual differences in cognitive demand influenced how effectively annotators used uncertainty visualization, we fit two LME models: one for annotation quality (mIoU) and one for editing time.  Each LME model includes treatment condition, self-reported cognitive load, and their interaction as fixed effects, with random intercepts for participants.

Self-reported cognitive load is similar across conditions (Uncertainty Visualization: $M = 3.02$, $SD = 1.33$; Baseline: $M = 3.23$, $SD = 1.12$), and this difference is not statistically significant ($p = 0.349$), indicating that we do not find evidence that the interface with uncertainty visualization systematically increases extraneous cognitive load.

The treatment main effect remains positive for annotation quality ($\beta =1.65$, $SE = 0.87$, $t = 1.89$, $p = 0.061$) and negative for editing time ($\beta = -3.74$, $SE = 3.12$, $t = -1.20$, $p = 0.233$) within these models, although attenuated relative to the primary analysis due to the inclusion of cognitive load as a covariate. Neither model shows a significant interaction between treatment condition and cognitive load (quality: $\beta = -0.32$, $SE = 0.26$, Wald $\chi^2(1) = 1.48$, $p = 0.224$; editing time: $\beta = 0.58$, $SE = 0.93$, Wald $\chi^2(1) = 0.39$, $p = 0.533$). As an exploratory analysis, we derive model-predicted quality estimates at $\pm 1~SD$ of cognitive load from the LME fixed effects. At low cognitive load (mean $- 1~SD$), the predicted treatment advantage is approximately $1.05$ mIoU points; at high cognitive load (mean $+ 1~SD$), this narrowed to approximately $0.28$ mIoU points. Although the interaction does not reach statistical significance, this directional pattern is consistent with the tool, which tends to be more effective for annotators with lower cognitive demand.

Finally, task satisfaction is descriptively higher in the uncertainty visualization group ($M = 4.05$, $SD = 1.25$) than in the baseline ($M = 3.88$, $SD = 1.13$), but this difference is not statistically significant (Mann-Whitney $U = 1{,}985$, $p = 0.331$, $r_b = -0.10$), ruling out the possibility that observed performance gains are driven by a general positive affect toward the interface rather than its functional utility.

\section*{Discussion}
\label{sec:discussion}
This work demonstrates that communicating localization uncertainty in the annotation interface can improve both label quality and annotation speed simultaneously, challenging the conventional speed vs.~accuracy trade-off in perceptual labeling tasks. 

\textbf{Mechanism: Guiding Attention toward Uncertain Predictions.}
The simultaneous improvement in quality and efficiency suggests that the uncertainty cues change not how much effort annotators put in, but where they allocate it. This interpretation is supported by box-level adjustment patterns as annotators in the uncertainty condition adjust high-uncertainty boxes at a higher rate, indicating that the color-coded signal is behaviorally effective in guiding inspection toward boxes with unreliable spatial boundaries. This interpretation is further supported by the difficulty-scaling pattern, in which effects are negligible on easy images and grow as image difficulty increases. Given that model predictions are already well-localized on easy images with low uncertainty, uncertainty cues carry little information beyond what the annotator can readily perceive, and benefits are accordingly absent. On more difficult images, where multiple boxes compete for attention~\cite{wolfe2021guided} and spatial errors are subtle, the cues reduce the effective visual search space and direct effort toward the boxes most in need of correction, consistent with classic accounts of guided visual search~\cite{itti2001computational}. We further find no statistically significant difference in cognitive load across conditions, indicating that the uncertainty interface does not alter the task's overall cognitive demand~\cite{sweller1988cognitive}. Together, these findings indicate that uncertainty communication through visual cues directs annotators to the relevant annotations rather than reducing the mental effort required to evaluate them.

The simultaneous gain in quality and efficiency is theoretically informative beyond its practical value. The conventional speed-accuracy trade-off in perceptual tasks rests on the assumption that quality and throughput compete for the same attentional resource, such that improving one typically costs the other~\cite{heitz2014speed}. This assumption holds when attention is correctly allocated by default. When it is not, as in unguided review, where annotators have no signal indicating where spatial errors are most likely, effort is distributed across boxes that vary widely in their need for correction. A signal that redirects this effort may therefore not increase attentional capacity, but reduce excessive burden. The simultaneous improvement in quality and speed is therefore precisely what an attentional redirection account predicts, i.e., time previously spent inspecting well-localized boxes is recovered and reallocated toward boxes that require correction, improving both outcomes at once.

\textbf{Leveraging Complementary Human–AI Capabilities.}
Both the baseline and treatment conditions represent a form of human–AI collaboration, but they differ in how effectively they exploit the complementary strengths each party brings. Detectors excel at processing large numbers of instances rapidly and consistently~\cite{rastogi2023taxonomy}, yet their spatial predictions degrade in precisely the conditions that human annotators can resolve, i.e., subtle spatial ambiguities that fall outside the model's reliable operating range. Without guidance, they must allocate this capacity broadly across all predictions, regardless of need. Uncertainty-guided annotation addresses this mismatch directly, aligning with the capability asymmetry between human and AI reviewers identified in human–AI collaboration research~\cite{hemmer2025complementarity}. The result is not simply a more accurate annotation pipeline, but a more efficient division of cognitive labor. This perspective also speaks to the ongoing debate between full automation and human-in-the-loop approaches. Rather than framing human-in-the-loop as a fallback when automation fails, uncertainty communication positions human reviewers as targeted contributors whose comparative advantage is actively mobilized. The value of keeping humans in the loop, therefore, depends not only on whether they are present, but also on whether they are directed toward decisions where their involvement genuinely matters.

\textbf{Implications for Annotation Workflows.}
The practical implications of these findings are relevant for model-assisted labeling workflows, since annotation errors on highly uncertain images, such as cluttered or occluded scenes, typically propagate into biased model training~\cite{song2022learning}. That this improvement in annotation quality comes with a concurrent gain in efficiency matters practically, since richer interfaces that slow annotators down often risk rejection in cost-sensitive production settings, hence making the concurrent efficiency gain in our study particularly relevant for real-world adoption. Both gains are focused where the model's spatial predictions are least reliable, which naturally points to combining our approach with active learning (AL), a strategy that may use the same uncertainty signal to prioritize which images are routed to human review in the first place~\cite{choi2021active,yang2024plug,sbeyti2025streamlining}. AL reduces the \textit{number of images requiring human attention}, while our interface improves how that \textit{attention is allocated within each image}, making the two mechanisms complementary without requiring additional model components. This integration, however, introduces a constraint as AL preferentially routes high-uncertainty images to human review~\cite{choi2021active}, meaning annotators will encounter these images disproportionately. Yet these are the images where uncertainty estimates are hardest to calibrate reliably, as calibration is known to degrade on the distributional tail precisely where uncertainty is highest~\cite{ovadia2019can}. Pipelines that combine AL selection with uncertainty-guided review should therefore ensure calibration of uncertainty estimates and continuously monitor them as the model is retrained on newly acquired labels, as we do in this work via isotonic regression.

\textbf{The Self-Reinforcing Cycle of Uncertainty-Guided Annotation.}
The benefits of localization uncertainty visualization may extend beyond any single annotation round through a self-reinforcing feedback loop across retraining cycles. Higher-quality annotations produced under uncertainty guidance yield a training dataset with higher fidelity, which in turn improves the detector's localization accuracy and the calibration of its uncertainty estimates. Better-calibrated uncertainty then produces more informative visual cues in the next round of human review, directing annotator attention more precisely toward the boxes that genuinely require correction. As the iterations proceed, uncertainty-guided annotation therefore becomes progressively more effective, creating a reinforcing cycle of better training datasets that inform more accurate models. Yet, in the early stages of training, when the models may be poorly calibrated and the uncertainty estimates may be inaccurate---consistent with the known degradation of calibration in spaces with few data and distributional shifts, as described in~\cite{ovadia2019can}---the cues may be less informative, or even misleading. In this case, the uncertainty information may add little or no value and may distract the human reviewer. Only as the model improves, uncertainty estimates sharpen, the guidance signal strengthens, and the efficiency gains documented here should widen. This temporal structure has a practical implication that the intervention is most valuable once an initial model has been trained on a seed dataset of sufficient quality, and its integration into the AL loop~\cite{choi2021active, yang2024plug}, where the same uncertainty signal governs both image routing and within-image attention allocation, provides the most favorable conditions for the cycle to compound.

\textbf{Automation Bias and the Risk of Underinspecting Low-Uncertainty Predictions.}
Our results demonstrate that uncertainty cues successfully redirect annotator effort toward high-uncertainty boxes, but the complementary risk deserves explicit attention. When a bounding box has low uncertainty, annotators may interpret this signal as a positive endorsement of correctness and reduce or skip inspection of that box entirely. This is a specific instance of automation bias~\cite{goddard2012automation}, the well-documented tendency for human operators in human-in-the-loop systems to overrely on outputs flagged as reliable and stop inspecting those that are not flagged for review. The box-level data in Figure~\ref{fig:pct-changed-albox} are consistent with this risk as participants in our treatment edit low-uncertainty boxes at a lower rate than those in the baseline, a pattern that is rational when uncertainty estimates are well-calibrated but that would constitute automation bias if they are not. Mitigating this bias requires rigorous calibration validation to detect potential biases, as well as interface designs that preserve a baseline inspection obligation even for low-uncertainty predictions, for example, by periodically surfacing low-uncertainty boxes for explicit review, or by monitoring correction rates on low-uncertainty boxes across annotator cohorts as an operational signal of calibration drift.

\textbf{Limitations.}
Several limitations warrant discussion. First, our work employs a between-subjects design with crowd-sourced participants on a single dataset (KITTI~\cite{geiger2012we}). While the KITTI benchmark is representative of autonomous driving scenarios and our participant pool was sufficiently powered to detect medium effect sizes, the generalizability to other domains (e.g., medical imaging~\cite{wang2021annotation}, satellite imagery~\cite{ahn2023human}) and annotation modalities (e.g., segmentation masks, keypoints) remains to be established. Second, the uncertainty estimates used in this study are derived from the model's own aleatoric uncertainty, meaning that the quality of the guidance is inherently bounded by the quality of the uncertainty calibration. Poorly calibrated models could direct attention to the wrong boxes, potentially degrading rather than enhancing annotation quality. Third, the treatment effect on the full sample attenuated to a non-significant trend when controlling for initial model prediction quality ($p = 0.136$), although it remained significant in the medium- and difficult subsets ($p = 0.040$). This suggests that the practical benefit is concentrated on non-trivial images. The intervention is most useful where it is most needed, but offers limited value for images that are already well-predicted. Fourth, the study duration was approximately 20 minutes per participant, which may not fully capture how annotators adapt to uncertainty cues over longer annotation sessions. Longitudinal studies examining learning effects and potential over-reliance on uncertainty cues would strengthen the evidence base.

\textbf{Future Work.}
Extending the uncertainty visualization to per-edge cues, i.e., highlighting which specific sides of a bounding box are uncertain rather than providing a single aggregate color, could result in a more granular attentional guidance. Integrating the approach with AL selection strategies would enable a unified uncertainty-driven pipeline from image selection to within-image annotation. Investigating adaptive interfaces that adjust how strongly uncertainty cues stand out based on annotator expertise or fatigue could further optimize the allocation of human attention. Finally, evaluating the downstream impact of uncertainty-guided annotations on detector retraining would close the loop between improvements in annotation quality and an increase in model performance, providing a complete picture of the value chain from uncertainty estimation to improved predictions.

In summary, this work identifies and addresses a gap in model-assisted annotation workflows where class uncertainty is often communicated to annotators via a score, whereas spatial uncertainty is not, thereby leaving humans reviewing those outputs without guidance on where localization errors are. By surfacing aleatoric localization uncertainty as a color-coded visual cue, we demonstrate that annotators can be systematically directed toward the predictions that most require correction. The simultaneous improvement in label quality and annotation speed, focused precisely on the boxes and images where model predictions are least reliable, points to a mechanism that guides attention to relevant instances while maintaining a similar level of overall cognitive demand. Together, these findings establish localization uncertainty as a practical and theoretically grounded lever for improving human-in-the-loop decision-making of spatial outputs. More broadly, the results suggest that uncertainty estimates should not remain internal to the model but be surfaced as actionable signals wherever human reviewers must verify outputs that are spatially precise, subtle to inspect, and consequential when wrong.

\section*{Method}
\label{sec:method}
We test our hypotheses in a randomized controlled experiment using a between-subjects design. We comply with all local ethical regulations. The research design is approved by the Institutional Review Board of our home university. All participants provide informed consent.

\subsection*{Dataset and Annotations}
\label{sec:dataset_annotations}

The KITTI dataset is a widely adopted autonomous driving dataset containing urban traffic scenes captured with a stereo camera system. We draw images from the KITTI validation split, which includes annotated bounding boxes for seven object classes: \textit{Car}, \textit{Van}, \textit{Truck}, \textit{Pedestrian}, \textit{Cyclist}, \textit{Tram}, and \textit{Misc}.

\textbf{Detection Model and Uncertainty Estimation.}
We obtain bounding box predictions and associated uncertainty estimates from a probabilistic Efficientdet-D0~\cite{tan2020efficientdet,mos-ks} pre-trained on MS COCO~\cite{lin2014microsoft} and fine-tuned on the KITTI~\cite{geiger2012we} training set for 200 epochs with 8 batches and an input resolution of 1024×512 pixels. All other hyperparameters of EfficientDet maintain their default value~\cite{tan2020efficientdet}. The model outputs, for each detected object, a bounding box $\mathbf{b} = [y_1, x_1, y_2, x_2]$, a detection confidence score, classification logits over the seven classes, and per-coordinate aleatoric localization uncertainty estimates. The raw per-coordinate uncertainties are calibrated using isotonic regression applied per class and per coordinate, and subsequently normalized by bounding box size to yield a relative uncertainty measure~\cite{kassem2023overcoming}. This normalization ensures that uncertainty values are comparable across objects of different scales. For visualization, the four per-coordinate uncertainties (top, left, bottom, right) are averaged into a single scalar per detection.

\textbf{Image Binning by Difficulty.}
To construct a balanced stimulus set spanning a range of annotation difficulties, we stratify images by their average localization uncertainty. For each image in the validation set, we computed the mean of the per-detection scalar uncertainties across all detections in the image. We then partition images into three equally sized bins using the 33rd and 67th percentiles of the image-level uncertainty distribution: \textit{Easy} (below the 33rd percentile), \textit{Medium} (between the 33rd and 67th percentiles), and \textit{Difficult} (above the 67th percentile). These bins serve as a proxy for image difficulty, with higher average uncertainty indicating scenes containing more difficult objects and faulty bounding boxes~\cite{kassem2023overcoming,sbeyti2025building}.

\textbf{Image Selection.}
Each participant is presented with 15 images, with five randomly sampled from each of the three difficulty bins and presented in randomized order to minimize ordering effects.

\textbf{Ground-truth Relabeling.}
The original KITTI annotations are known to contain missing and imprecise bounding boxes, particularly for highly occluded or distant objects~\cite{northcutt2021confident,sbeyti2025building}. To obtain a reliable gold standard against which to evaluate participant annotations, three members of the author team independently re-annotated all 97 images used in the experiment. This relabeling effort corrected spatial imprecision in existing bounding boxes and added annotations for objects that were present in the scene but absent from the original ground-truth. As reported in Supplementary Table~S4, the original 566 bounding boxes were expanded to 955, with 430 new annotations added and 41 erroneous labels removed. The majority of additions (98.8\%) were \textit{Car} instances missed due to heavy occlusion, small apparent size, or partial visibility at image boundaries (Supplementary Table~S5a). These corrected annotations serve as the ground-truth for all metrics reported in this work. They are released at \url{https://github.com/mos-ks/MUHA/tree/main/data/relabeled_ground_truth/labels} and can be interactively inspected at \url{https://mos-ks.github.io/MUHA/}.

\subsection*{Procedure}
\label{sec:procedure}
The experiment procedure is as follows. First, participants receive information about the study objectives and compensation. After giving informed consent, each participant is randomly assigned to one of two conditions: a baseline condition ($n = 60$) or an uncertainty condition ($n = 60$).

The study begins with a familiarization phase in which participants read instructions explaining the annotation task, the interface, and the compensation structure. Participants are required to use a desktop PC or laptop to ensure optimal visual functionality. Annotated screenshots then introduce the annotation interface and the specific behaviors expected of annotators (see Supplementary Figure~S1), i.e., adjusting the position and size of pre-generated bounding boxes to maximize spatial accuracy. In the uncertainty condition, additional instructional material explains the color-coded uncertainty cues (a continuous color-coded bar, with red indicating high uncertainty and blue indicating low uncertainty) and how to interpret them.

Following the tutorial, participants complete a comprehension check consisting of multiple-choice questions covering core task concepts, including how to handle overlapping objects, the meaning of localization uncertainty, and which bounding boxes to verify. Participants who answer incorrectly were given one opportunity to retry.

Before the main task, participants complete pre-task questionnaires assessing self-efficacy and task familiarity, each measured on a 6-point Likert scale (ranging from \emph{Strongly Disagree} to \emph{Strongly Agree} and \emph{Not Familiar} to \emph{Extremely Familiar}, respectively). One item within the self-efficacy scale serves as an embedded attention check, asking participants to select a specific response.

Participants then complete a pre-test consisting of two practice images to verify that they have understood the task and can successfully execute it. In the first image, the goal is to familiarize the participants with the tool. For the second image, they are required to improve the mIoU from an initial value of 87.5\% to at least 90\% within two attempts to ensure sufficient engagement with the task. Failure to achieve these results results in exclusion from the study.

In the main task, each participant reviews and corrects bounding box annotations for the 15 driving-scene images drawn from KITTI. Participants can zoom in and out of each image and were instructed to adjust box positions and sizes to fit each object as tightly as possible. Object classifications are pre-fixed and cannot be modified. The study is designed to take approximately 20 minutes.

After completing the annotation task, participants fill out a post-task survey measuring cognitive load, intention to use, task satisfaction, and perceived helpfulness of the uncertainty visualization. A full list of the survey items is in Supplementary Table~S1. One item in the task difficulty scale serves as a second attention check. Participants also respond to three open-ended questions asking them to reflect on how the uncertainty visualization influences their annotation behavior and prioritization decisions.

\subsection*{Annotation Interface and Conditions}
\label{sec:conditions}

The annotation interface is developed using Streamlit~\cite{streamlit2026} and delivered to participants via a web browser. Participants are randomly assigned to one of two conditions, which differ solely in whether localization uncertainty information is displayed within the annotation interface:

\begin{enumerate}
    \item \emph{Baseline} presents pre-annotated images with standard bounding boxes around detected objects. No uncertainty information is provided. Participants can freely adjust box positions and sizes. This reflects the standard model-assisted annotation workflow, in which annotators review and correct model-generated proposals without additional guidance.

    \item \emph{Uncertainty Visualization} augments the same pre-annotated images with color-coded bounding box borders reflecting the model's localization uncertainty. Red borders indicate high uncertainty (i.e., the model is less confident about the box's spatial position), while blue borders indicate low uncertainty. Participants are instructed to pay particular attention to high-uncertainty boxes, while still verifying all annotations.
\end{enumerate}

\subsection*{Outcome}
\label{sec:outcome}

To evaluate the effect of uncertainty visualization, the primary outcome is \emph{annotation quality}, measured as the mIoU between participant-adjusted bounding boxes and ground-truth labels, averaged across all pre-annotated objects within each image. IoU is defined as the ratio of the intersection over the union of the predicted and ground-truth boxes, hence indicating spatial accuracy.

The secondary outcome is \emph{annotation efficiency}, measured as the time in seconds participants spent reviewing and adjusting each image.

We also collect additional measures through a post-task survey to assess perceived interactions with the uncertainty visualization (e.g., helpfulness, task satisfaction, and cognitive load). Self-reported task familiarity is collected prior to the main task and used as covariates in the statistical models.

\subsection*{Study Participants}

The study is conducted online using the crowdsourcing platform Prolific, selecting the participant pool for ``AI task'' and specifying the task as ``annotation''. Participants are required to be fluent in English and reside in the United Kingdom or the United States. No prior annotation experience is required. Participants receive a base payment of £2.50, with a performance-based bonus of up to £1.00 determined by their annotation accuracy, which is intended to incentivize high-quality annotations while preserving ecological validity.

Overall, we exclude participants who (1) fail more than one attention check across the study, (2) report technical problems, or (3) do not reach an mIoU of at least 90\% during the pre-test within two attempts, indicating an inability to perform the task as intended. A total of 120 participants meet the inclusion criteria and form the final sample for analysis, with 60 participants assigned to each condition.

\subsection*{Statistical Analysis}

To test the effect of uncertainty visualization on annotation quality and efficiency, we first conduct one-tailed independent-samples Welch $t$-tests comparing the two conditions on participant-level means of mIoU and annotation time, with significance level $\alpha = 0.05$. Effect sizes are reported as Cohen's $d$. 

To examine whether treatment benefits vary as a function of image difficulty, we stratify performance across the three levels (Easy, Medium, Difficult) and conduct one-tailed $t$-tests within each level, aggregating mIoU to participant level within each bin prior to testing.

To assess whether individual differences in cognitive load moderate treatment efficacy, we fit two additional LME models: one for annotation quality and one for editing time. Each model includes treatment condition, self-reported cognitive load, and their interaction as fixed effects, and random intercepts for participants. The overall model fit is evaluated via likelihood ratio chi-square tests against intercept-only null models, and interaction terms are assessed using Wald chi-square tests. As a descriptive exploratory analysis, we derive model-predicted quality estimates at $\pm 1~SD$ of cognitive load from the fitted LME fixed effects. Task satisfaction between conditions is compared using a two-tailed Mann-Whitney $U$ test with rank-biserial correlation as the effect size measure.

All analyses are conducted in \textsf{R} (version~4.4.1) using the packages \texttt{tidyverse}, \texttt{lme4}, \texttt{lmerTest}, and \texttt{coin}.

% -----------------------------------------------------------------------------
\subsection*{Robustness Checks}
% -----------------------------------------------------------------------------
 
We conduct a series of robustness checks to ensure the validity of our findings, with full results reported in Supplementary Tables~S1--S2. (a)~To mitigate the influence of extreme observations, we winsorize mIoU values at the 5th and 95th percentiles. (b)~We re-fit the primary LME model incorporating self-efficacy and task familiarity as participant-level covariates, to rule out the possibility that pre-existing individual differences between groups drive the effect. (c)~We estimate an LME model including the initial quality of labels generated through our model as a covariate to control for image-level difficulty independently of the bin stratification. This check is applied to both the full sample and the medium$+$difficult secondary analysis subset; the latter constitutes the theoretically relevant test, as on easy images, the covariate is expected to absorb variance where uncertainty cues provide no benefit. Across checks (a) and (b), the findings remain fully consistent with the main results. For check (c), the treatment effect survives in the medium$+$difficult subset ($p = 0.040$) but attenuates to a non-significant trend in the full sample ($p = 0.136$), consistent with the theoretical prediction that uncertainty cues add value only on non-trivial images.

\backmatter
\section*{Acknowledgments}
This work has been partially funded by the Pilot Program for Core Informatics (COIN) at the KIT of the Helmholtz Association. It was also supported by the Helmholtz Association Initiative and Networking Fund on the HAICORE@KIT partition.

% \section*{References}
% ...
% \end{thebibliography}

% \bibliography{sn-bibliography}% common bib file

\clearpage

% =========================================================
% Supplementary Information starts here
% =========================================================

% Reset figure, table, section, equation counters so that
% supplementary items are numbered S1, S2, ... independently
% of the main paper.
\setcounter{figure}{0}
\setcounter{table}{0}
\setcounter{section}{0}
\setcounter{equation}{0}

\renewcommand{\thefigure}{S\arabic{figure}}
\renewcommand{\thetable}{S\arabic{table}}
\renewcommand{\thesection}{S\arabic{section}}
\renewcommand{\theequation}{S\arabic{equation}}

\section*{Supplementary Material}

\section{Questionnaire Details and Interface Tutorial}
\begin{table}[ht]
\centering
\caption{Survey items used in the pre- and post-task questionnaires. All items were measured on a 6-point Likert scale.}
\label{supptab:items}
\renewcommand{\arraystretch}{1.15}
\begin{tabular}{@{}lp{0.50\textwidth}p{0.27\textwidth}@{}}
\toprule
\textbf{Construct} & \textbf{Item} & \textbf{Scale} \\
\midrule
\multicolumn{3}{l}{\textit{Pre-task measures}} \\[0.3em]
\multirow{3}{*}{Self-Efficacy} & I am confident about my ability to do the task. & \multirow{3}{4cm}{1 = Strongly Disagree, 6 = Strongly Agree} \\
 & I am self-assured about my capabilities to perform the task. & \\
 & I have mastered the skills necessary for the task. & \\[0.5em]
\multirow{2}{*}{Task Familiarity} & How familiar are you with image annotation in general? & \multirow{2}{4cm}{1 = Not Familiar, 6 = Extremely Familiar} \\
 & How familiar are you with image annotation in the field of autonomous driving? & \\[0.5em]
\midrule
\multicolumn{3}{l}{\textit{Post-task measures}} \\[0.3em]
\multirow{3}{*}{Cognitive Load} & During this task, it was difficult to recognize and link the crucial information. & \multirow{3}{4cm}{1 = Strongly Disagree, 6 = Strongly Agree} \\
 & During this task, it was exhausting to find the important information. & \\
 & The design of the task was very inconvenient for learning. & \\[0.5em]
\multirow{3}{*}{Intention to Use} & I intend to continue using provided AI support when it is offered to me. & \multirow{3}{4cm}{1 = Strongly Disagree, 6 = Strongly Agree} \\
 & Using provided AI support for future tasks is something I would do. & \\
 & I predict that I would use provided AI support in the future when applicable. & \\[0.5em]
\multirow{3}{*}{Task Satisfaction} & Overall, how satisfied are you with your performance on this task? & \multirow{3}{4cm}{1 = Very Little, 6 = Very Much} \\
 & Overall, how much did you learn? & \\
 & Overall, how much did you enjoy performing this task? & \\[0.5em]
\multirow{1}{*}{Helpfulness} & The uncertainty visualization was helpful in making my subsequent decisions. & 1 = Strongly Disagree, 6 = Strongly Agree \\
\bottomrule
\end{tabular}
\end{table}
% \clearpage
%%%%%  Screenshots
\begin{figure}[p]
    \centering
    \captionsetup[subfigure]{font=small} 
    % Row 1: Introduction
    \begin{subfigure}{0.48\textwidth}
        \centering
        \includegraphics[width=\linewidth]{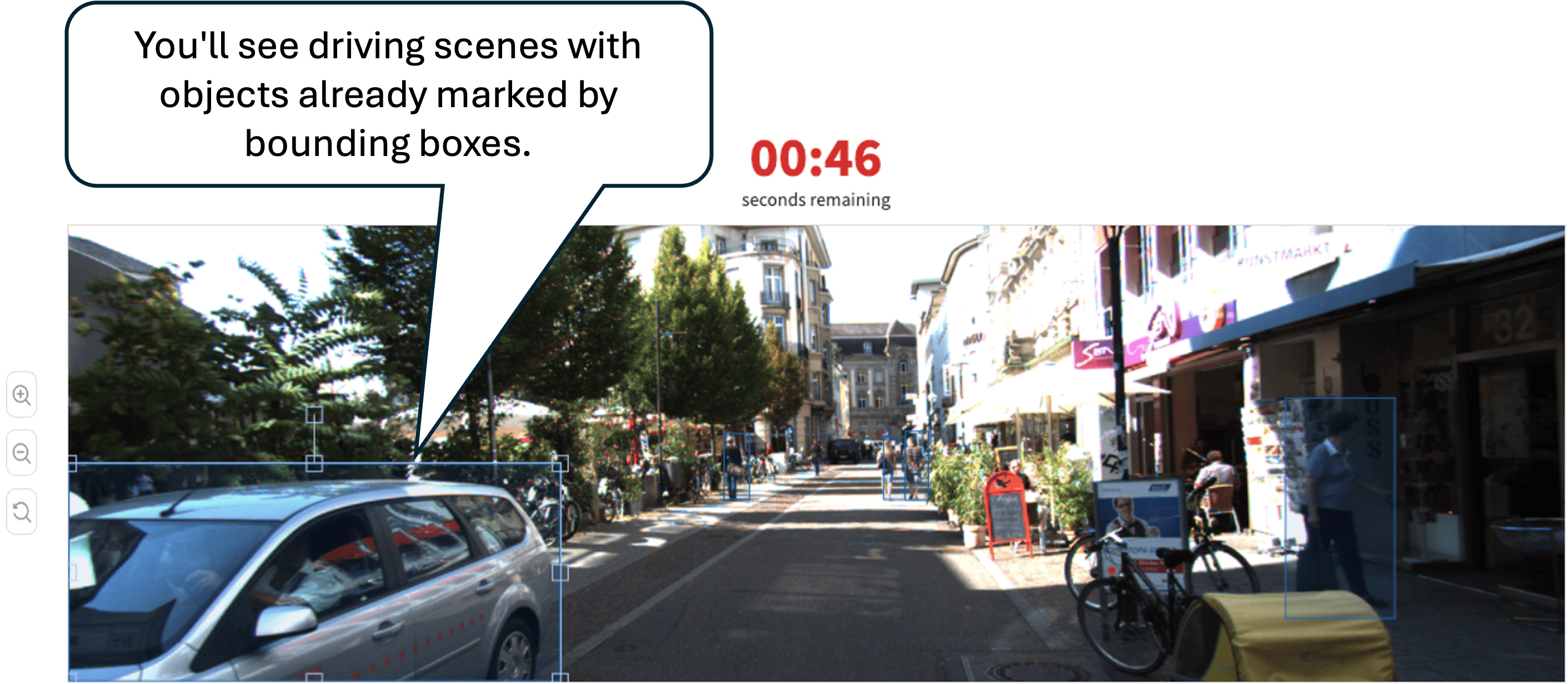}
        \caption{Baseline: Introduction}
    \end{subfigure}
    \hfill
    \begin{subfigure}{0.48\textwidth}
        \centering
        \includegraphics[width=\linewidth]{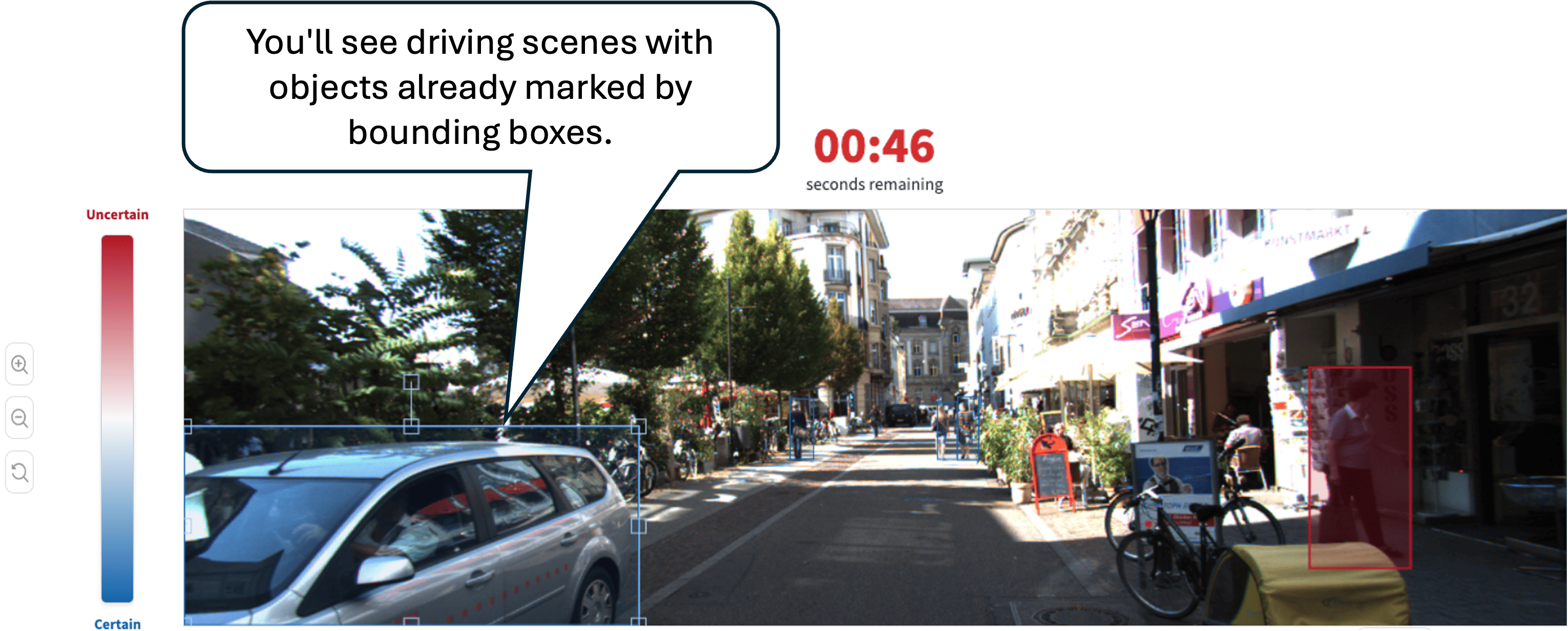}
        \caption{Treatment: Introduction}
    \end{subfigure}

    % \vspace{1em}

    % Row 2: Interaction
    \begin{subfigure}{0.48\textwidth}
        \centering
        \includegraphics[width=\linewidth]{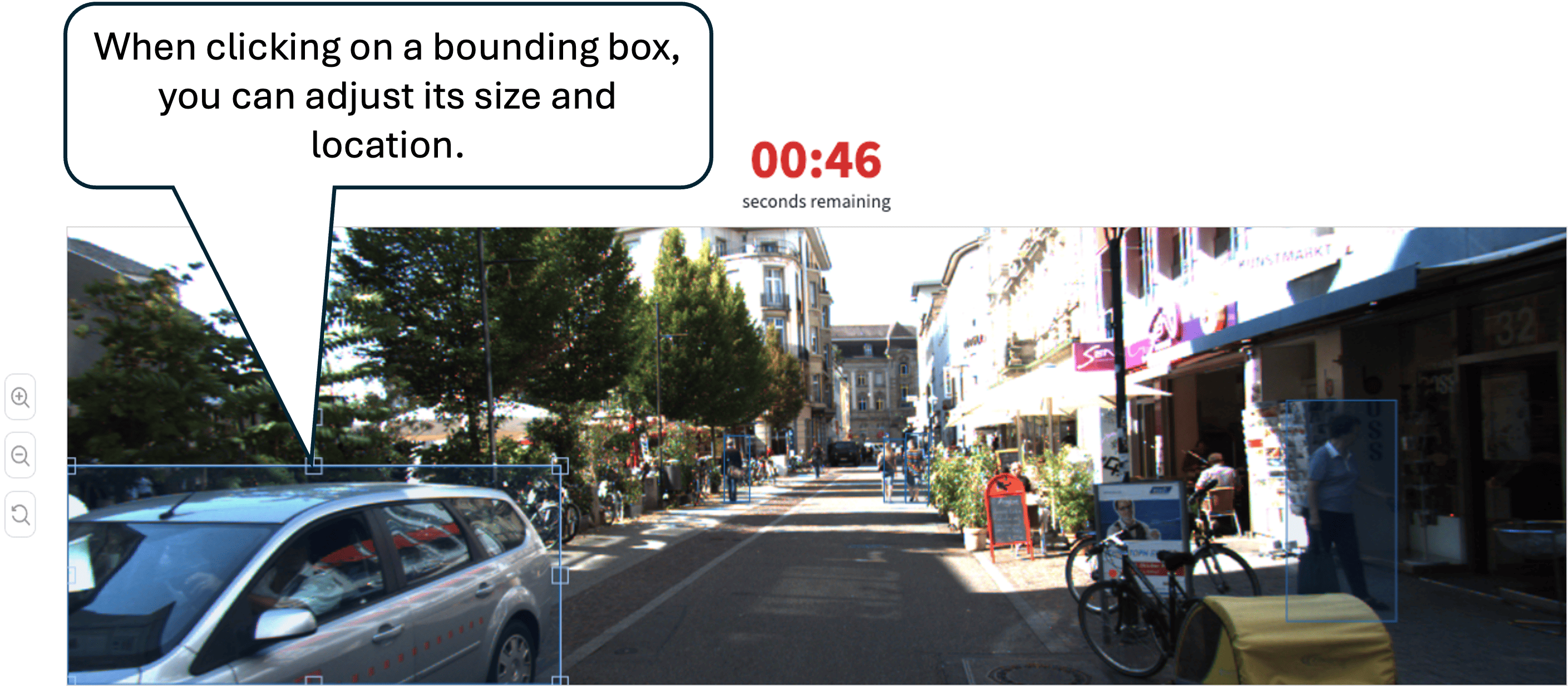}
        \caption{Baseline: Bounding box adjustment}
    \end{subfigure}
    \hfill
    \begin{subfigure}{0.48\textwidth}
        \centering
        \includegraphics[width=\linewidth]{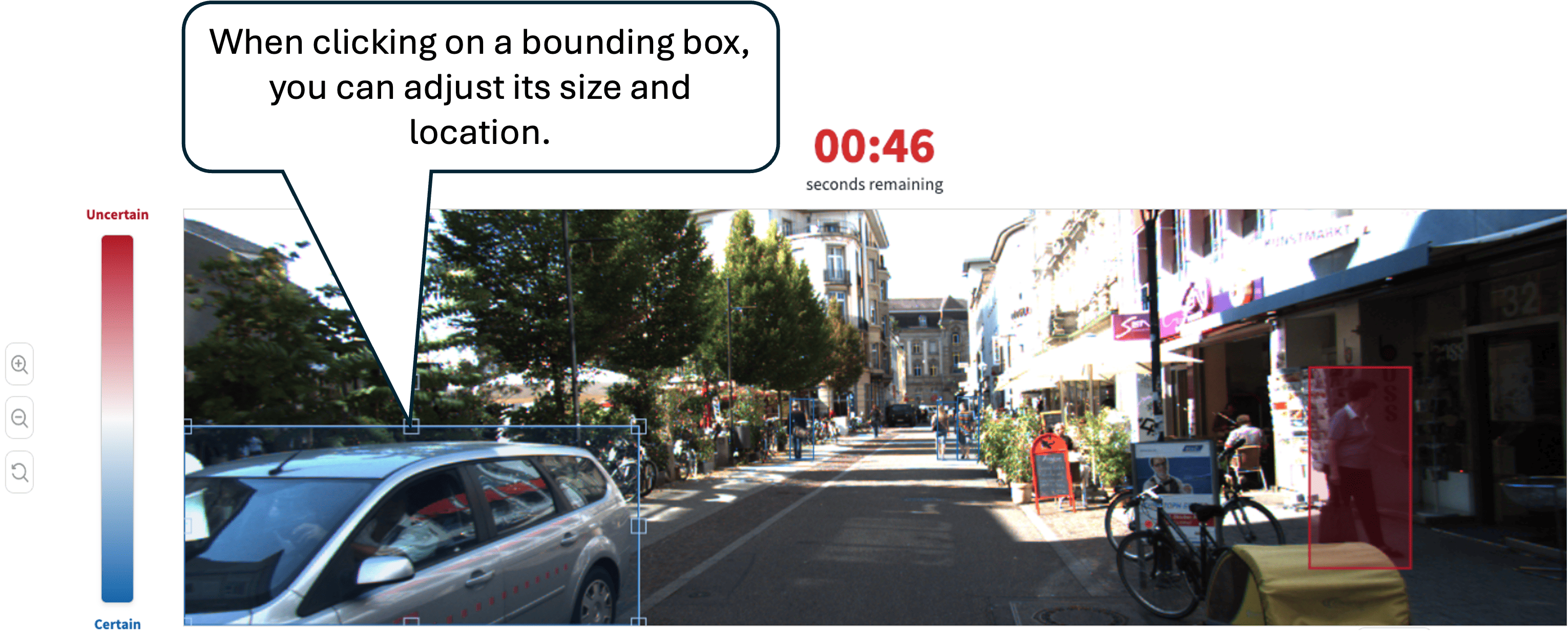}
        \caption{Treatment: Bounding box adjustment}
    \end{subfigure}

    % \vspace{1em}

    % Row 3: Timer
    \begin{subfigure}{0.48\textwidth}
        \centering
        \includegraphics[width=\linewidth]{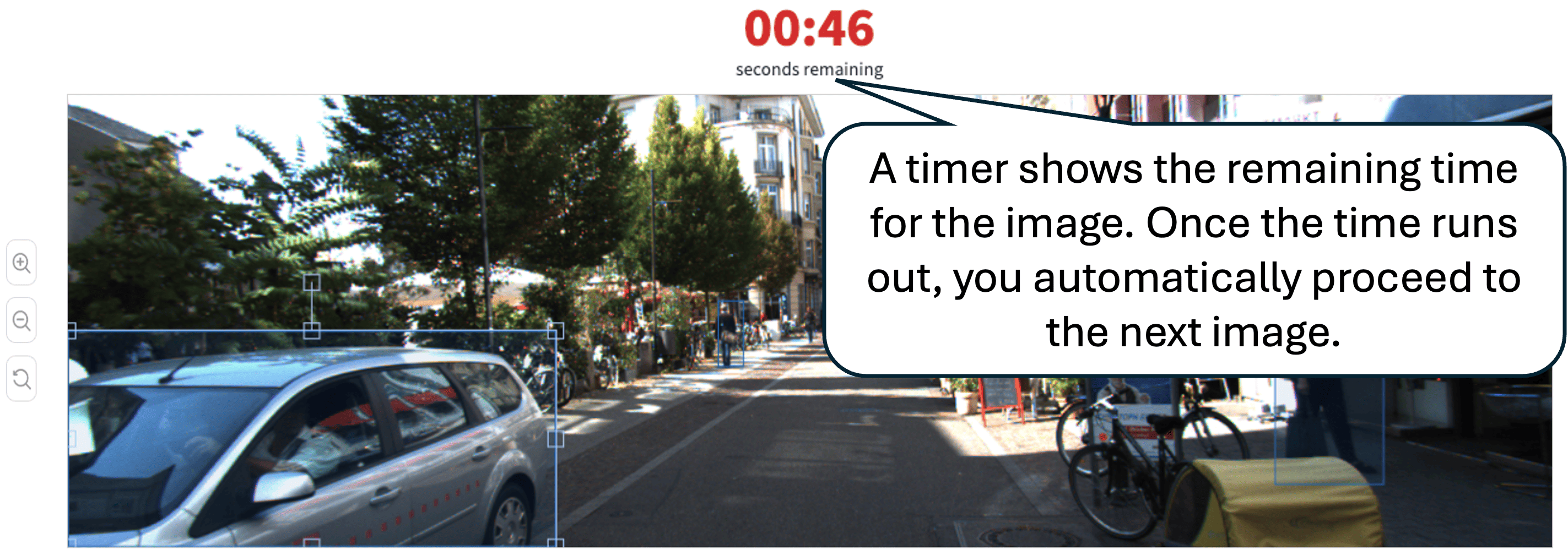}
        \caption{Baseline: Task timer}
    \end{subfigure}
    \hfill
    \begin{subfigure}{0.48\textwidth}
        \centering
        \includegraphics[width=\linewidth]{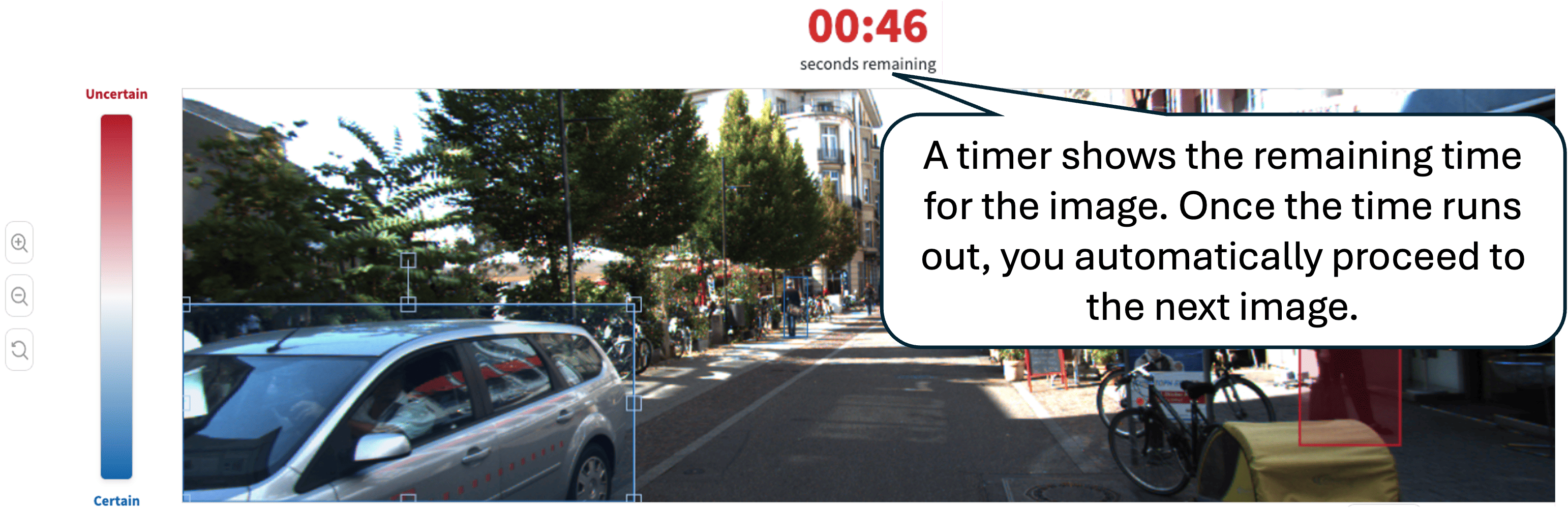}
        \caption{Treatment: Task timer}
    \end{subfigure}

    % \vspace{1em}

    % Row 4: Controls (Zoom)
    \begin{subfigure}{0.48\textwidth}
        \centering
        \includegraphics[width=\linewidth]{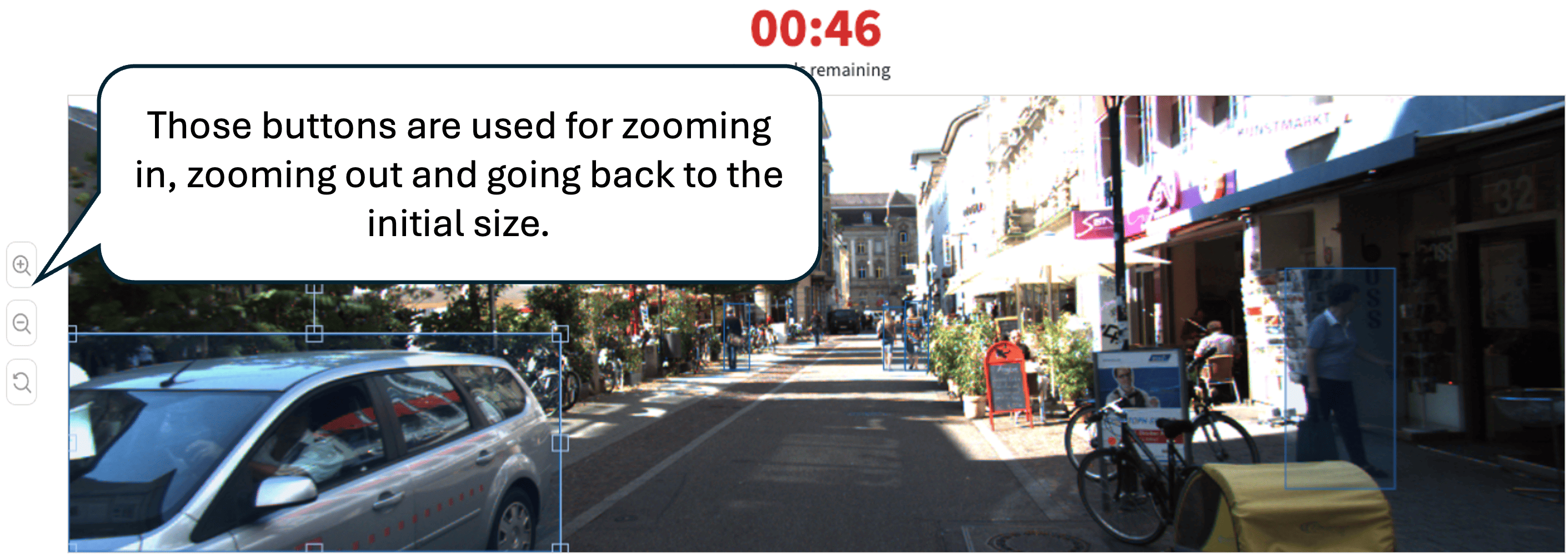}
        \caption{Baseline: Zoom controls}
    \end{subfigure}
    \hfill
    \begin{subfigure}{0.48\textwidth}
        \centering
        \includegraphics[width=\linewidth]{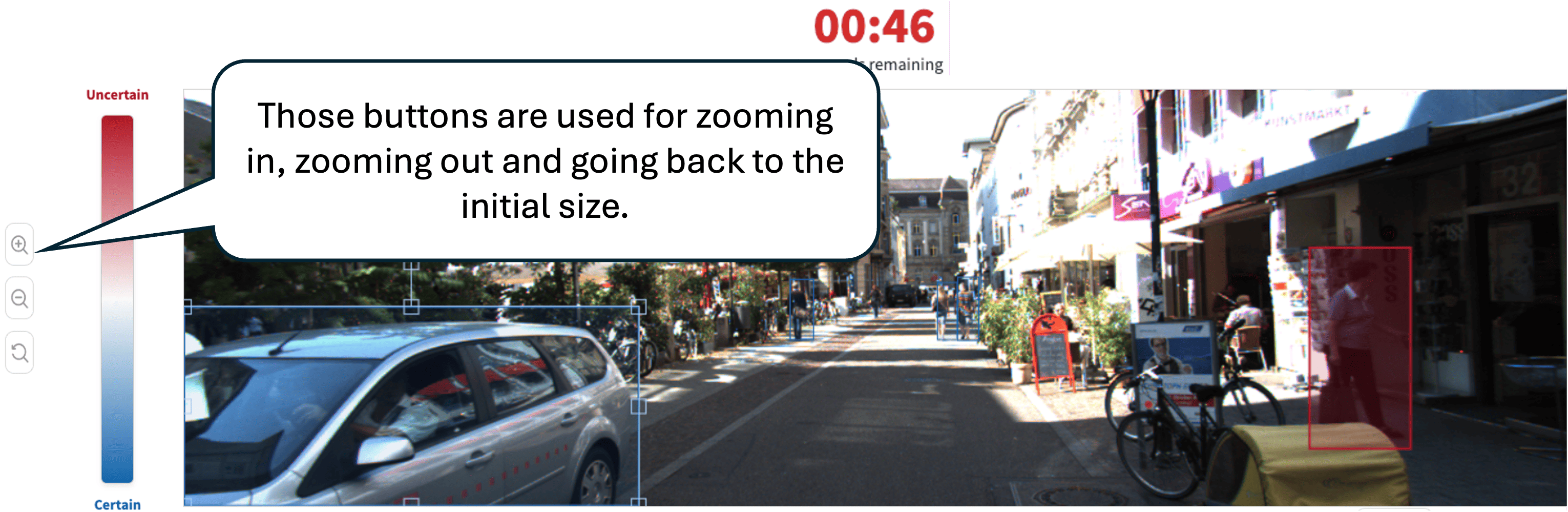}
        \caption{Treatment: Zoom controls}
    \end{subfigure}

    % \vspace{1em}

    % Row 5: Reload
    \begin{subfigure}{0.48\textwidth}
        \centering
        \includegraphics[width=\linewidth]{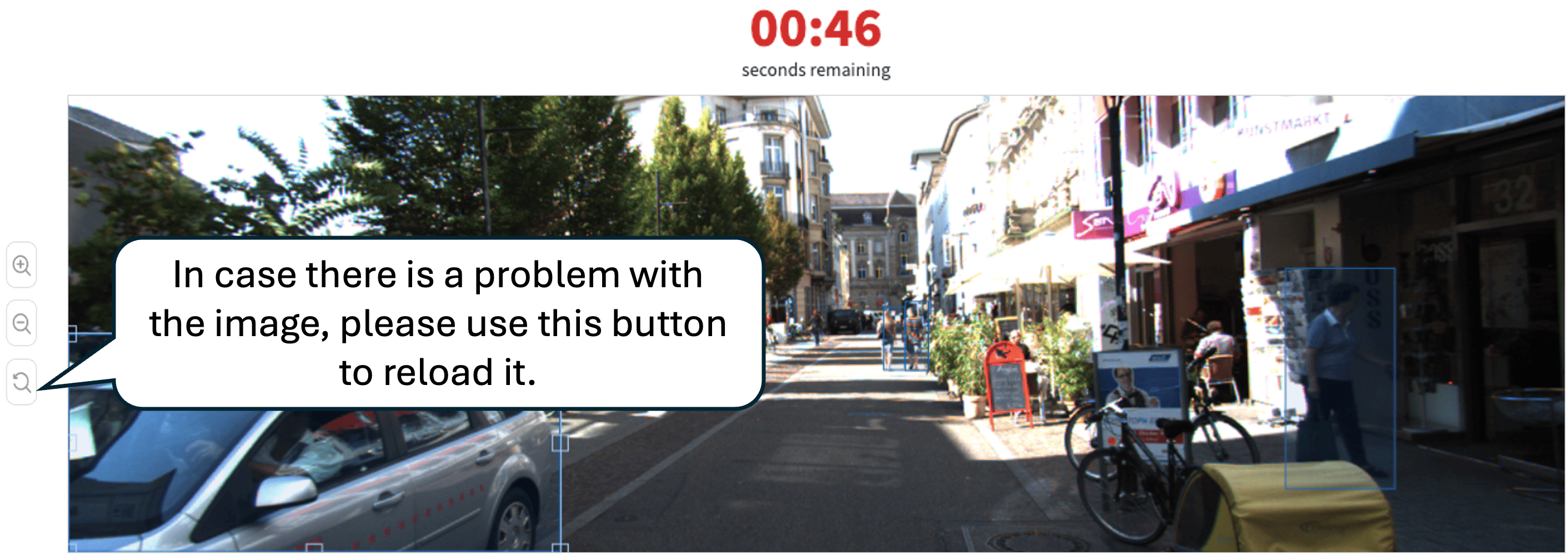}
        \caption{Baseline: Reload functionality}
    \end{subfigure}
    \hfill
    \begin{subfigure}{0.48\textwidth}
        \centering
        \includegraphics[width=\linewidth]{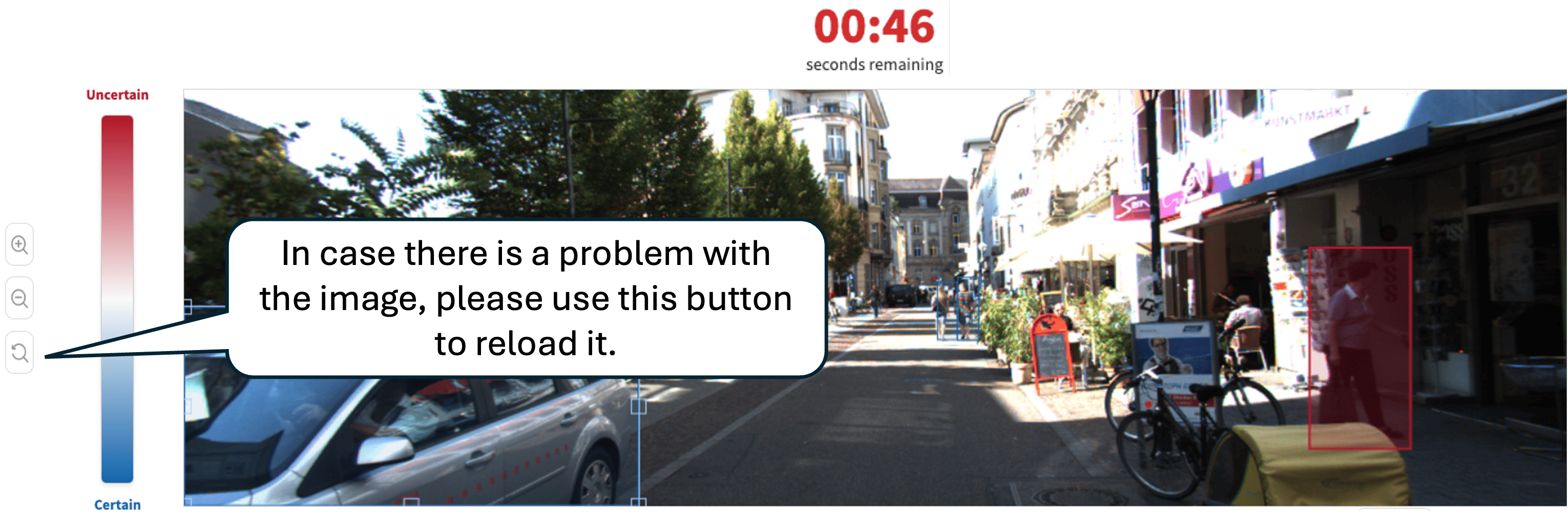}
        \caption{Treatment: Reload functionality}
    \end{subfigure}

    % \vspace{1em}

    \begin{subfigure}{\textwidth} % Width added here
        \centering
        \includegraphics[width=0.5\linewidth]{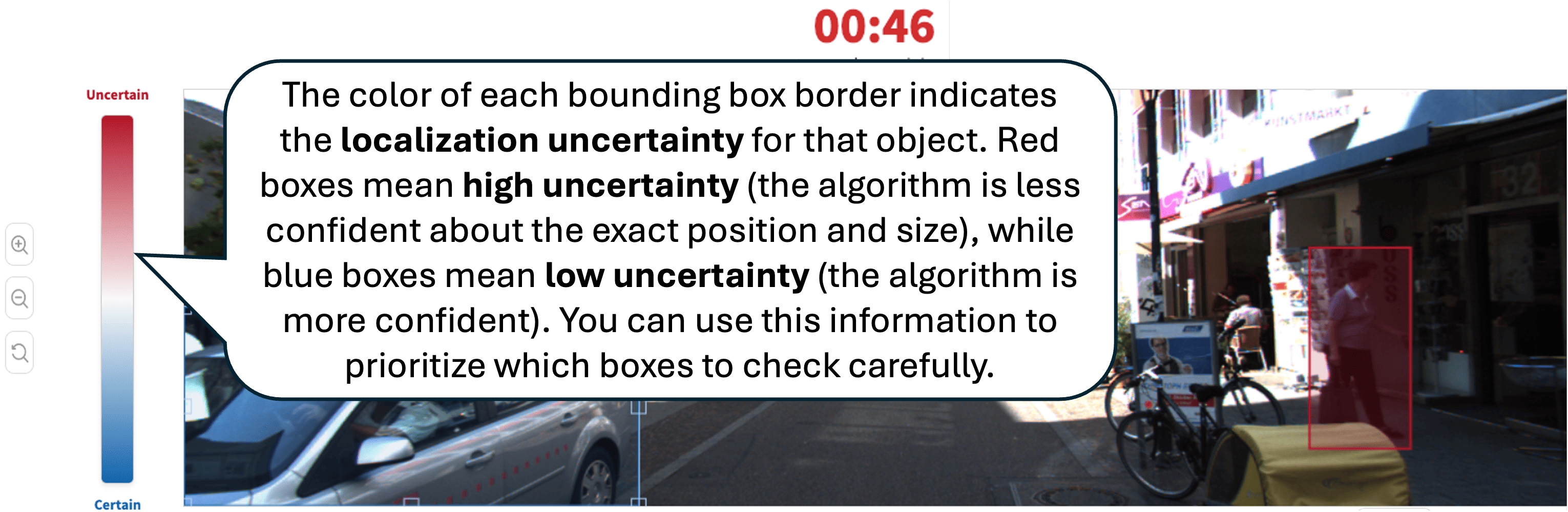}
        \caption{Treatment: Explanation of Uncertainty}
    \end{subfigure}

    \caption{\textbf{Interface Tutorial and Condition Comparison.} Overview of the instructional slides presented to participants in the Baseline (\textit{left}) and Uncertainty Visualization (\textit{right}) conditions. The treatment condition differs by the inclusion of the uncertainty color scale (Certain/Blue to Uncertain/Red) and the corresponding color-coding of bounding box borders.}
    \label{fig:supp_interface_tutorial}
\end{figure}

\clearpage
\section{Robustness Checks with Full Sample}

\begin{table}[ht]
\centering
\begin{minipage}{\textwidth}
\centering
\caption{Robustness checks for the primary treatment effect on annotation quality (mIoU), full sample ($N = 120$ participants, $1{,}800$ trials). All models include a random intercept for participant and use ML estimation. The primary LME result is included for reference. Robustness checks (a, b, and c) follow the same order and notion as introduced in the robustness checks subsection.}
\label{tab:robustness_full}
\vspace{0.5em}
\begin{tabular*}{\textwidth}{@{\extracolsep{\fill}}lcccc}
\toprule
Model & $\hat{\beta}$ (treatment) & $SE$ & $t$ & $p$ \\
\midrule
Primary LME (treatment only)        & 0.698 & 0.317 & 2.203 & 0.030 \\
a: Winsorized mIoU (5th--95th pct) & 0.678 & 0.291 & 2.335 & 0.021 \\
b: $+$ self-efficacy, familiarity  & 0.707 & 0.315 & 2.249 & 0.026 \\
c: $+$ initial model mIoU          & 0.403 & 0.268 & 1.502 & 0.136\footnotemark[1] \\
\bottomrule
\end{tabular*}

\vspace{0.5em}
\begin{minipage}{\textwidth}
\footnotesize
\footnotemark[1] The attenuation in c is expected as the initial mIoU absorbs variance in easy images where uncertainty cues offer low or no effect for focusing attention. The treatment effect in this check is therefore conservative by design. The mid$+$high secondary analysis (Supplementary Table S2) addresses this directly.
\end{minipage}
\end{minipage}
\end{table}

\section{Robustness Checks (Medium$+$Difficult Difficulty)}

\begin{table}[ht]
\centering
\begin{minipage}{\textwidth}
\centering
\caption{Primary results and robustness checks for the secondary analysis restricted to medium and difficult images ($N = 120$ participants, $1{,}200$ trials, 10 images each). Random intercept for participant, ML estimation throughout. Robustness checks (a,b, and c) follow the same order and notion as introduced in the robustness checks subsection.}
\label{tab:robustness_subset}
\vspace{0.5em}
\begin{tabular*}{\textwidth}{@{\extracolsep{\fill}}lcccc}
\toprule
Outcome & $t$ & $df$ & $p$ (one-tailed) & $d$ \\
\midrule
mIoU    &  2.686 & 117.9 & 0.004 &  0.490 \\
Editing time  & $-$2.038 & 117.7 & 0.022 & $-$0.372 \\
\bottomrule
\end{tabular*}
\vspace{0.3em}
\raggedright \textbf{Panel A: Primary $t$-tests}
\vspace{1em}\\
\raggedright \textbf{Panel B: LME robustness checks (outcome: mIoU)}
\vspace{0.3em}
\centering
\begin{tabular*}{\textwidth}{@{\extracolsep{\fill}}lcccc}
\toprule
Model & $\hat{\beta}$ (treatment) & $SE$ & $t$ & $p$ \\
\midrule
Primary LME (treatment only)        & 0.870 & 0.321 & 2.709 & 0.008 \\
a: Winsorized mIoU (5th--95th pct) & 0.837 & 0.291 & 2.881 & 0.005 \\
b: $+$ self-efficacy, familiarity  & 0.881 & 0.317 & 2.777 & 0.006 \\
c: $+$ initial model mIoU          & 0.553 & 0.266 & 2.077 & 0.040 \\
\bottomrule
\end{tabular*}

\vspace{0.5em}
\begin{minipage}{\textwidth}
\footnotesize
\textit{Note:} Unlike the full-sample result, the treatment effect survives control for initial model mIoU within the mid$+$high subset ($p = 0.040$), confirming that the benefit on harder images is not a proxy for differences in starting annotation quality.
\end{minipage}
\end{minipage}
\end{table}

\clearpage
\section{Ground-truth Relabeling}
\begin{wraptable}{r}{0.37\textwidth}
\centering
\caption{Summary of ground-truth relabeling corrections.}
\label{tab:relabel_summary}
    \resizebox{0.37\textwidth}{!}{%
\begin{tabular}{lr}
\toprule
Metric & Value \\
\midrule
Images corrected & 97 \\
Original bounding boxes & 566 \\
Corrected bounding boxes & 955 \\
Retained (IoU $> 0.5$) & 525 \\
Newly added annotations & 430 \\
Removed annotations & 41 \\
Net change & $+389$ \\
Avg.\ changes per image & 4.9 \\
\bottomrule
\end{tabular}%
    }%
\end{wraptable}We manually inspect and correct the ground-truth annotations for 97 images from the KITTI validation set. Table~\ref{tab:relabel_summary} summarizes the overall relabeling effort: out of 566 original bounding boxes, 525 are retained (IoU $> 0.5$), 430 new annotations are added, and 41 erroneous or ambiguous labels are removed, yielding a net increase of 389 bounding boxes (68.7\%).

Table~\ref{tab:relabel_perclass} reports the per-class distribution of changes. The vast majority of additions correspond to \textit{Car} instances (425 out of 430, 98.8\%), which are predominantly missing in the original annotations due to heavy occlusion, small apparent size at large distances, or partial visibility at image boundaries. In addition, 494 existing bounding boxes are modified to better align with the visible object extent. The highest modification rates are observed for \textit{Car} (363), \textit{Pedestrian} (42), and \textit{Van} (38), reflecting systematic inaccuracies in the original tight-fitting boxes. Removed annotations primarily belong to the \textit{Misc} (11) and \textit{Tram} (7) categories, which are often mislabeled or marked inconsistently in the original ground-truth.
\begin{table}[ht]
\centering
\begin{subtable}{0.43\textwidth}
    \centering
    \resizebox{\textwidth}{!}{%
    \begin{tabular}{lrrr}
    \toprule
    Class & Added & Modified & Removed \\
    \midrule
    Van & 0 & 38 & 1 \\
    Car & 425 & 363 & 13 \\
    Misc & 0 & 4 & 11 \\
    Truck & 0 & 17 & 1 \\
    Pedestrian & 3 & 42 & 6 \\
    Tram & 0 & 5 & 7 \\
    Cyclist & 2 & 25 & 2 \\
    \midrule
    Total & 430 & 494 & 41 \\
    \bottomrule
    \end{tabular}%
    }%
    \caption{Per-class distribution of relabeling changes.}
    \label{tab:relabel_perclass}
\end{subtable}%
\hfill
\begin{subtable}{0.52\textwidth}
    \centering
    \resizebox{\textwidth}{!}{%
    \begin{tabular}{lrrrr}
    \toprule
    Image ID & Original & Corrected & Added & Removed \\
    \midrule
    000497 & 9 & 24 & 15 & 0 \\
    000101 & 3 & 17 & 14 & 0 \\
    000361 & 7 & 21 & 14 & 0 \\
    000243 & 5 & 17 & 12 & 0 \\
    000359 & 3 & 13 & 10 & 0 \\
    000594 & 4 & 14 & 10 & 0 \\
    000601 & 5 & 15 & 10 & 0 \\
    000674 & 8 & 14 & 8 & 2 \\
    000743 & 5 & 15 & 10 & 0 \\
    000961 & 7 & 15 & 9 & 1 \\
    \bottomrule
    \end{tabular}%
    }%
    \caption{Top 10 images with the most annotation changes.}
    \label{tab:relabel_topimages}
\end{subtable}
\caption{Granular ground-truth relabeling statistics across the seven KITTI object categories and top 10 images.}
\end{table}

\begin{figure}
    \centering
    \includegraphics[width=1\linewidth]{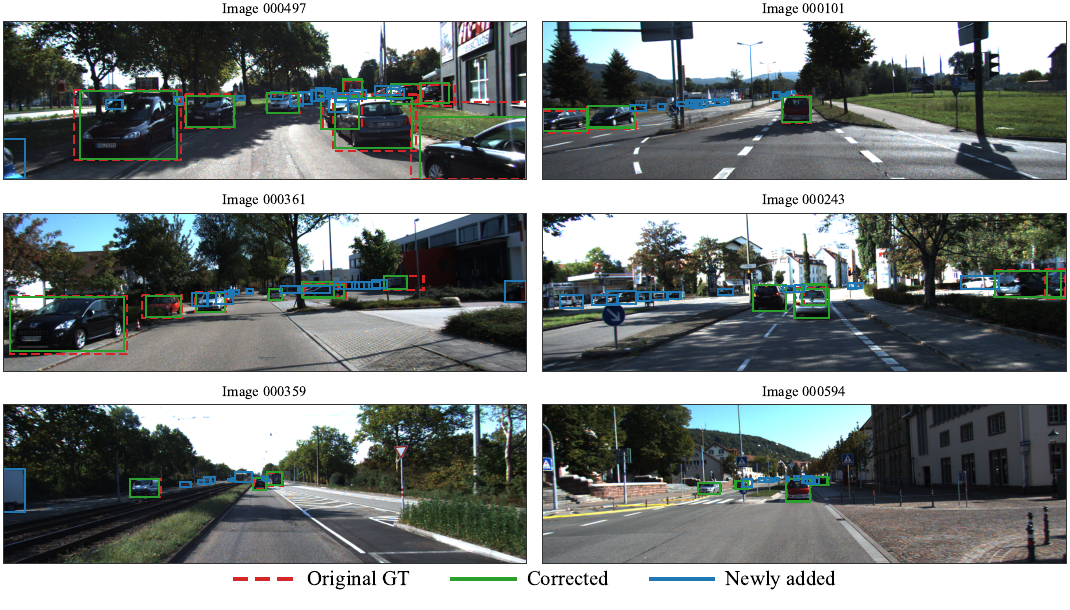}
    \caption{Visualization of six representative KITTI examples~\cite{geiger2012we}, showing the original annotations (red, dashed), corrected annotations (green, solid), and newly added bounding boxes (blue, solid).
}
    \label{fig:relabel_comparison}
\end{figure}

Table~\ref{tab:relabel_topimages} lists the ten images with the most annotation changes. These scenes are characterized by dense traffic with numerous partially occluded vehicles, where up to 15 bounding boxes are added to a single frame (e.g., image 000497). Fig.~\ref{fig:relabel_comparison} visualizes six representative examples, showing the original annotations (red, dashed), corrected annotations (green, solid), and newly added bounding boxes (blue, solid).

\end{document}